\documentclass[%
 reprint,
 twocolumn,
 amsmath,amssymb,
 aps,
prl,
]{revtex4-2}

\usepackage[dvipdfmx]{graphicx}% Include figure files
\usepackage{dcolumn}% Align table columns on decimal point
\usepackage{bm}% bold math
\usepackage{color}
\usepackage{braket}
\usepackage{lipsum}
\usepackage{mathdots}
\usepackage[version=3]{mhchem} 
\usepackage{ulem}
\usepackage{color, soul}
\usepackage{xspace}
\usepackage{url}
\newcommand{\TMA}  {\ce{Ti2MnAl}\xspace}
\newcommand  {\eqn}[1]{(\ref{eqn:#1})}

\renewcommand{\_}[1]  {_\textrm{#1}} 

\preprint{APS/123-QED}

\begin{document}

\title{Topological spin...}% Force line breaks with \\

\date{\today}% It is always \today, today,

\title{
Topological Spin-Orbit Torque in Ferrimagnetic Weyl Semimetal
}

\author{Tomonari Meguro$^1$}
 \email{meguro.tomonari@mbp.phys.kyushu-u.ac.jp} 
\author{Akihiro Ozawa$^{2}$}
 \email{akihiroozawa@issp.u-tokyo.ac.jp} 
\author{Koji Kobayashi$^{3}$}%\thanks{k-koji@sophia.ac.jp}}
 \email{k-koji@sophia.ac.jp}
\author{Yasufumi Araki$^{4}$}%\thanks{araki.yasufumi@jaea.go.jp}}
 \email{araki.yasufumi@jaea.go.jp}
\author{Kentaro Nomura$^1$}%\thanks{nomura.kentaro@phys.kyushu-u.ac.jp}}
 \email{nomura.kentaro@mbp.phys.kyushu-u.ac.jp}

 \affiliation{%
$^1$Department of Physics, Kyushu University, Fukuoka 819-0395, Japan
}%
 \affiliation{%
$^2$Institute for Solid State Physics, The University of Tokyo, Kashiwa 277-8581, Japan
}%
 \affiliation{%
$^3$Physics Division, Sophia University, Chiyoda-ku, Tokyo 102-8554, Japan
}%
\affiliation{%
$^4$Advanced Science Research Center, Japan Atomic Energy Agency, Tokai, Ibaraki 319-1195, Japan}

\begin{abstract}
The spin-orbit torque~(SOT) in a compensated ferrimagnetic Weyl semimetal, \TMA, is studied by the linear response theory.
We elucidate that the SOT driven by all the occupied electronic states is present in magnetic Weyl semimetal, unlike in conventional metallic magnets.
Around the energy of the Weyl points, we find that such an SOT is dominant and almost independent of the disorder.
The emergence of the SOT in \TMA can be understood from
the structure of the mixed Berry curvature around the Weyl points, which is similar to that of the ordinary Berry curvature.
\end{abstract}

 \maketitle

%------ 1st paragraph ---------------------------------------------------------------------------------------------------------------
{\it Introduction---}%
Efficient electrical control of electron spins~\cite{Dyakonov1971, Kato2004, Sinova2015, Manchon2019} is one of the fundamental aims of spintronics.
Spin-orbit coupling~(SOC) plays a significant role by correlating the electron spin and orbital motion.
The SOC enables the electrical generation of spin polarization in heterostructures~\cite{Sinova2015, Miron2010, Miron2011, Suzuki2011, Emori2013, Jamali2013, Garello2013} or bulk noncentrosymmetric crystals~\cite{Zelezny2014, Zelezny2017,Wadley2016}.
The generated spin polarization acts as an effective magnetic field for magnetic moments,
whose effect is called the spin-orbit torque~(SOT).
The SOT has been intensely studied as a promising method for functional control of magnetization~\cite{chernyshov2009,Miron2010,liu2012current,liu2012spin,fukami2016}.
However, to obtain a large SOT, one needs to inject a large longitudinal current,
which suffers from the energy dissipation by the Joule heating.
Such an SOT is termed the ``dissipative SOT''.
The problem of dissipation is crucial in metals with a large density of states~(DOS) near the Fermi level $E\_F$,
where longitudinal conductivity $\sigma_{xx}$ is proportional to the DOS.
The dissipation comes from the scattering of a large number of electrons around $E\_F$ by disorders and phonons.
To avoid the dissipation, we need systems 
that
can generate the SOT with a small longitudinal current.

%------ 2nd paragraph ---------------------------------------------------------------------------------------------------------------
To suppress the dissipation, %effect,
we focus on the
characteristic transport properties of magnetic Weyl semimetals (MWSMs).
The MWSMs feature linear dispersion and point nodes called Weyl points~(WPs)
~\cite{Murakami2007, Wan2011, Burkov2011} .
Since the DOS near the energy of the WPs is negligibly small, $\sigma_{xx}$ tends to be suppressed, leading to 
the small dissipation
in the longitudinal transport.
The SOT in bulk MWSMs has been studied 
based on the continuum model with 
isotropic spin-momentum locking and broken inversion symmetry~\cite{DKurebayashi2021, Burkov2023}.
The SOT obtained in these studies
needs a finite DOS by shifting $E\_F$ away from the WPs, similar to $\sigma_{xx}$, corresponding to the dissipative SOT. 
By contrast, 
non-dissipative
transport phenomena, such as the anomalous Hall effect,
arise from the nontrivial band topology in MWSMs.
The band topology is characterized by the Berry curvature 
(BC) 
from the WPs in momentum space~\cite{Armitage2018, Xiao2010}. 
The anomalous Hall conductivity $\sigma_{yx}$ corresponds to the integral of the BC over all the occupied electronic states below $E\_F$.
Such a mechanism is termed the ``topological effect''.
% 

%------ 3rd paragraph ---------------------------------------------------------------------------------------------------------------
% In contrast to dissipative SOT, 
The existence of the SOT driven by all the occupied electronic states below $E\_F$ has been pointed out in several metallic magnets~\cite{HKurebayashi2014,Zelezny2014, Zelezny2017,Wadley2016}.
We here refer to such a non-dissipative SOT
as the ``topological SOT''.
%materials other than MWSMs.
Theoretically, the topological
SOT can be characterized by a mixed Berry curvature~(MBC)~\cite{Freimuth2014}.
The MBC is the extension of the BC in 
momentum space to the composite parameter space 
spanned by momentum and magnetization.
In the systems analyzed so far, 
the magnitude of topological SOT is
smaller than the dissipative SOT.
To enhance the topological SOT with low Joule heating, 
materials showing a large MBC are desired.
% 

%------ 4th paragraph ---------------------------------------------------------------------------------------------------------------
%
In this letter, we find a large topological SOT in a model of MWSM \TMA.
\TMA is one of the candidate systems of ideal MWSMs with compensated ferrimagnetic ordering~\cite{Shi2018}.
In \TMA, both the time-reversal and inversion symmetries are broken.
We show that, in \TMA, the topological SOT is significantly larger than dissipative SOT near the energy of WPs with a typical disorder strength.
Compared with the SOTs studied in the bulk of noncentrosymmetric antiferromagnets \cite{Zelezny2014,Zelezny2017,Wadley2016},
we find a giant
effective magnetic field
in \TMA.
These properties enable us to 
electrically control the compensated ferrimagnetic ordering 
with low dissipation in the bulk system.
By analyzing the distribution of the MBC,
we find that the origin of the large topological SOT is the MBC enhanced around the WPs.
With such a large topological SOT,
we expect the potential utility of noncentrosymmetric MWSMs, including \TMA, for integrated spintronics devices.
%-------------------------------------------------------------------------------------------------------------------------------------

\begin{figure}[t]
   \includegraphics[width=1.0\hsize]{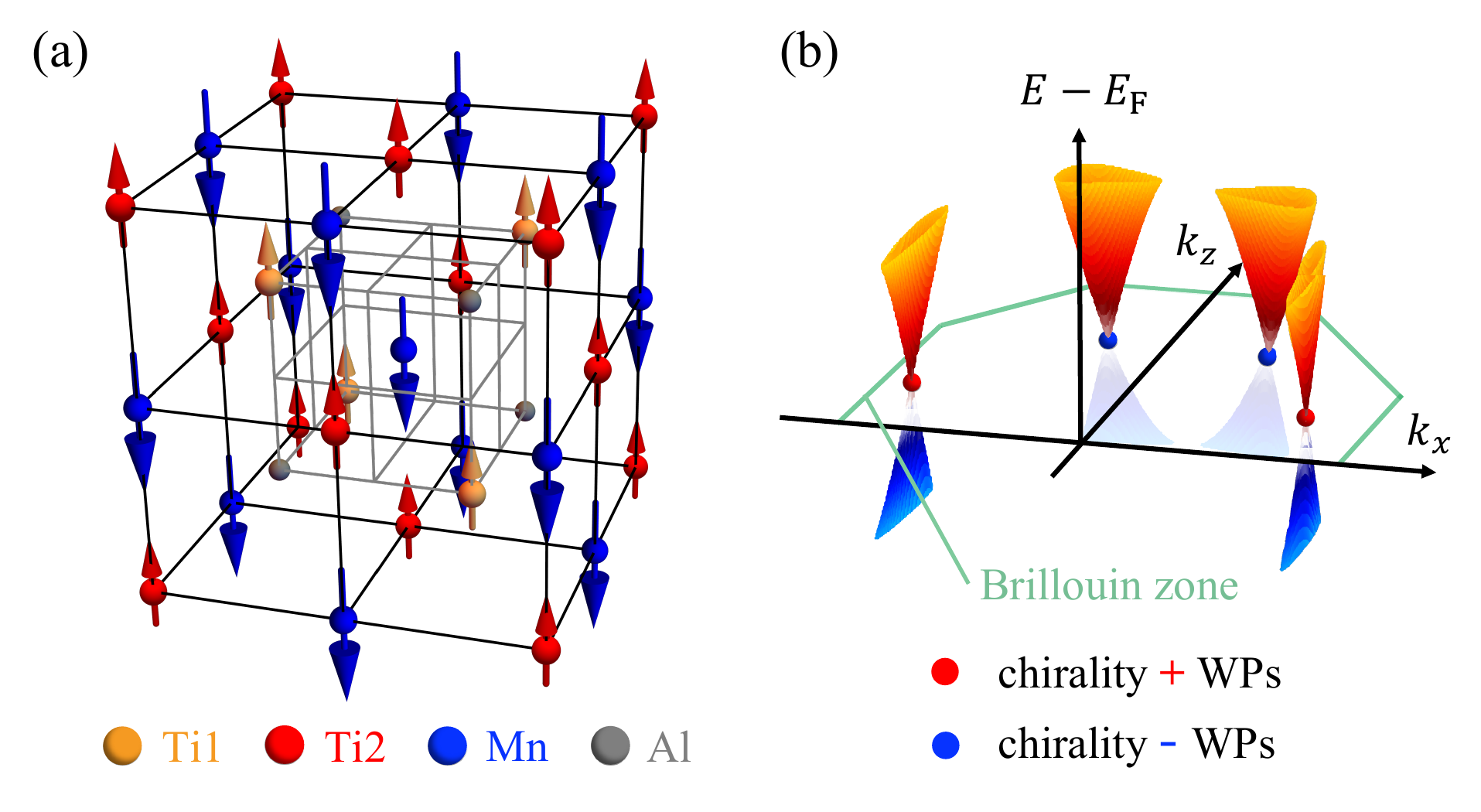}
\caption{
(a)~Crystal structure of \TMA.
Ti and Mn are responsible for the compensated ferrimagnetic ordering.
%% The green, light blue and purple solid lines represent inter-sublattice nearest-neighbor hoppings.
% The blue solid line represents intra-sublattice nearest-neighbor hoppings.
% Note that only some hopping is shown for visibility.
(b)~The energy dispersion around the WPs on $k_{y}=0$ plane.
} 
\label{fig:lattice}
\end{figure}

{\it Model---}
% \textcolor{blue}{
Before proceeding to our analysis, first
% }
we explain the electronic properties of the ferrimagnetic Weyl semimetal \TMA~\cite{Shi2018}.
The crystal structure of \TMA is an inverse Heusler type with space group ${F}{\bar 4}3m$~\cite{Skaftouros2013, Feng2015}  as shown in Fig.~\ref{fig:lattice}(a).
Magnetic moments at two Ti sites anti-parallelly align to that at the Mn site, showing zero net magnetization.
The first-principles calculations of the electronic band structure predict
that the WPs are located near $E\_F$, where the DOS is negligibly small.
These properties lead to the large anomalous Hall angle $\theta_{\rm AHE}=\sigma_{yx}/\sigma_{xx}$~\cite{Shi2018}.
Since not only the time-reversal symmetry but also the inversion symmetry is broken due to its crystalline structure,
the SOT is permitted in bulk.

%-----------------------------------------------%

%----------------- paragraph 2 ------------------% ここではTi2MnAlの結晶構造-->磁性の話
For our calculations, we introduce a tight-binding model of \TMA~\cite{Meguro2024}.
In reality, all $d$ orbitals mainly contribute to the electronic structure near $E\_F$. 
For our minimal model to describe the low-energy band structure around $E\_F$,
we consider a single isotropic orbital for two Ti and Mn as the basis to satisfy the symmetry of the crystalline structure.
Here, to distinguish two different Ti sites,
we use the notations Ti1 and Ti2 as shown in Fig.~\ref{fig:lattice}(a).
We neglect the orbital at the Al site for simplicity,
while the effect of Al is substantially incorporated via SOC as discussed later.
% 
%-----------------------------------------------%
%===== FIG =====%
% ===== FIG =====%
%----------------- paragraph 3 ------------------%ここでは全ハミルトニアンを出す．基底（軌道）の紹介．バンド構造の観点からAl軌道は無視する．
Our Hamiltonian $H$ consists of a hopping term, on-site energy, SOC, and exchange coupling:
%$H$ is given by
%
\begin{align}\label{eqn:Ham0}
 H = 
   &  -\!\sum_{\braket{ij}\alpha \beta} \sum_{s}
         t_{\alpha \beta} C^{\dagger}_{i \alpha s} C_{j \beta s}
      +\sum_{i \alpha} \sum_{s}
         \epsilon_{\alpha} C^{\dagger}_{i \alpha s} C_{i \alpha s} \nonumber\\
   &  +\!i\frac{2\lambda_{\rm SOC}}{a^2} \sum_{\braket{ij} \alpha} \sum_{s s'} 
         C^{\dagger}_{i \alpha s} 
         \left[ ( \bm{d}^{\alpha ij}_{1} \times \bm{d}^{\alpha ij}_{2} ) \cdot \bm{\sigma} \right]_{ss'}
         C_{j \alpha s'} \nonumber\\
   &  -\!\sum_{i \alpha} \sum_{s s'} J_{\alpha} 
         C^{\dagger}_{i \alpha s} 
         ( \bm{\sigma} \cdot \hat{\bm{n}} )_{s s'} 
         C_{i \alpha s'}.
\end{align}
Here, $C^{\dagger}_{i \alpha s}$ denotes an electron creation operator of 
% sublattice $\alpha =$ A~(Ti1), B~(Ti2), and C~(Mn) with the spin $s = \uparrow, \downarrow$ at $i$ site.
sublattice $\alpha =$ Ti1, Ti2, and Mn with the spin $s = \uparrow, \downarrow$ at $i$ site.
We briefly explain each term in the following
~(see Appendix A for details of our tight-binding model).
%-----------------------------------------------%

%----------------- paragraph 3 ------------------%  ここではNN，NNN hoppingの話．fig1.(a)中の色つき棒を参照しながら紹介していく．
% 
The first term includes the inter- and intra-sublattice hoppings.
The second term denotes the on-site energy $\epsilon_\alpha$ for each sublattice. 
%-----------------------------------------------%
%----------------- paragraph 4 ------------------%  ここでスピン軌道相互作用
The third term 
% \textcolor{blue}{
describes the effect of
% }
SOC due to the local breaking of inversion symmetry 
% \textcolor{blue}{
for the intra-sublattice hoppings.
% } 
% The dominant asymmetry comes from the imbalance between Ti1 and Al sites. 
$\bm{d}^{\alpha ij}_{1,2}$ are the two nearest-neighbor hopping vectors from the site $i$ to $j$ of the sublattice $\alpha$,
which are introduced according to the Fu-Kane-Mele model~\cite{Fu2007}. 
We consider SOC for the Ti2-Ti2 and Mn-Mn hoppings with the same strength $\lambda_{\rm SOC}$.
We neglect the SOC for the Ti1-Ti1 
% \textcolor{magenta}{
for simplicity.
% \\
%----------------- paragraph 5 ------------------%ここで交換相互作用
The fourth term represents the exchange coupling between conduction electron spin $\bm{\sigma}$ and
the ferrimagnetic ordering.
The direction of the ferrimagnetic ordering is represented by the unit vector $\hat{\bm{n}}$, and is fixed to the $z$-direction in the following calculations.
% magnetic moment ${\hat\bm{n}}$.
%$\bm{s}_{{A},i} = a^{\dagger}_{is} \(\bm{\sigma}\)_{ss'} a_{is'}$ is the spin operator of the conduction electron at A site, 
%and the same for  $\bm{s}\_{B}$ and $\bm{s}\_{C}$.
% $\bm{\sigma}$ is the Pauli matrix vector representing electron spin.
% 
The ferrimagnetism is encodeded in the mean-field coupling strength $J_\alpha$ for each sublattice $\alpha$,
where the signs of $J_{\rm Ti1}$ and $J_{\rm Ti2}$ are taken opposite to $J_{\rm Mn}$.
%-----------------------------------------------%

%----------------- paragraph 6 ------------------%  ここで，para.2で出したハミルトニアンから得られる電子状態についておさらい．またフェルミ準位の設定

%========================================
% \textcolor{blue}{
Once the model is given,
% }
we recall the electronic structure 
% \textcolor{blue}{
around $E\_F$.
% }
The energy spectrum obtained by the Bloch Hamiltonian of $H$, denoted as $\mathcal{H}(\bm{k})$ 
exhibits the WPs as shown in Fig.~\ref{fig:lattice}(b).
Red and blue points correspond to the WPs with positive ($+$) and negative ($-$) chiralities, respectively.
We obtain eight WPs on the  $k_{x} = 0$ plane, 
% \textcolor{blue}{
as well as
% } 
on the $k_{y} = 0$ and $k_{z} = 0$ planes,
and thus 24 WPs in total.
%
% For example, on the $k_{y} = 0$ plane,
% the linear dispersions around the WPs are formed as shown in Fig.~\ref{fig:lattice}(b). 
We determine $E\_F$ from the 4/6 filling condition.
In this study, we focus on the behavior of SOT around the energy of the WPs, $-E_0 \le E\_F \le E_0$, where $E_0=50$~meV 
% \textcolor{blue}{
characteries the energy scale exhibiting the Weyl dispersion around the WPs.
% }

% 
{\it Electrically induced spin density---}
% \textcolor{blue}{
We now proceed to the analysis of the SOT in this model.
% }
The conduction electron spin driven by the electric field acts as an effective magnetic field 
and exerts the SOT on the magnetic moments at $\ce{Ti}1$, $\ce{Ti}2$, and $\ce{Mn}$.
% 
% To study the effective magnetic field,
% % in this section, 
% we first study the electrically induced spin density at each sublattice,
% based on the above tight-binding model.
% 
In the linear response regime, the spin density at sublattice $\alpha$ ($= {\rm Ti1}, {\rm Ti2}, {\rm Mn}$) is calculated by using the Kubo formula~\cite{Manchon2019}:
%-- eq --%
\begin{align}
  \frac{\braket{\sigma^{\alpha}_{\mu}}}{E_{x}} 
  &= i\hbar \lim_{\bm{q}\rightarrow 0} \sum_{n \neq m}
      \int_{\bm{k}} 
      \frac{ f_{n\bm{k}} - f_{m\bm{k}+\bm{q}} }
           { E_{n\bm{k}} - E_{m\bm{k}+\bm{q}} } 
      \frac{S^{\alpha,\mu}_{nm,\bm{k}+\bm{q}} J^{x}_{mn,\bm{k}+\bm{q}} }
           {E_{n\bm{k}} - E_{m\bm{k}+\bm{q}} + i\eta}
      % \frac{ \langle n\bm{k}|  \frac{\hbar}{2}\hat{\sigma}^{\alpha}_{\mu} |m{\bm{k} \rangle} 
      % \langle m\bm{k}| (-e \hat{v}_{x })         |n{\bm{k} \rangle} }
      % {E_{n\bm{k}} - E_{m\bm{k}} + i\eta}
.
\label{eqn:SpinDens}
\end{align}
%-- eq --%
%
Here, the index $\mu(=x,y,z)$ denotes the direction 
% \textcolor{blue}{
of the induced spin density,
% }
whereas the direction of the external electric field 
% \textcolor{blue}{
$\boldsymbol{E}$ is fixed to the $x$-direction.
% }
% 
We define the spin matrix element
$ S^{\alpha, \mu}_{nm,\bm{k}+\bm{q}} = \langle n\bm{k}|  \hat{\sigma}^{\alpha}_{\mu} |m \bm{k}+\bm{q} \rangle $,
where $\hat{\sigma}^{\alpha}_{\mu}$ is the electron spin operator at the sublattice $\alpha$,
and $|n{\bm{k} \rangle}$ is the eigenstate with eigenenergy $E_{n\bm{k}}$ of $\mathcal{H}(\bm{k})$.
$ J^{x}_{mn,\bm{k}+\bm{q}} = \langle m\bm{k}+\bm{q}| (-e \hat{v}_{x}) |n{\bm{k} \rangle}$ is the current matrix element where $-e \hat{\bm{v}} =-\frac{e}{\hbar}\frac{\partial \mathcal{H}}{\partial \bm{k}}$ is the current operator.
$f_{n\bm{k}} = 1/(e^{(E_{n\bm{k}}-E\_F)/{k\_{B}T}}+1)$ is the Fermi-Dirac distribution function,
% \textcolor{blue}{
with $T=0$ here.
% }
%
% \textcolor{blue}{
The integral is taken over the Brillouin zone, with $\int_{\bm{k}} = (2\pi)^{-3} \int d^3\bm{k}$.
% }
For the momentum integration, the momentum space is split into the mesh of $250\times 250\times 250$ for all calculations in this paper.
We introduce the impurity effect by the first Born approximation, 
so that the self energy part $\eta$ in Eq.~\eqn{SpinDens} is replaced by $\frac{\hbar}{2\tau}$ with the relaxation time $\tau$.
%

%===== FIG =====%
\begin{figure}[t]
   \includegraphics[width=1.0\hsize]{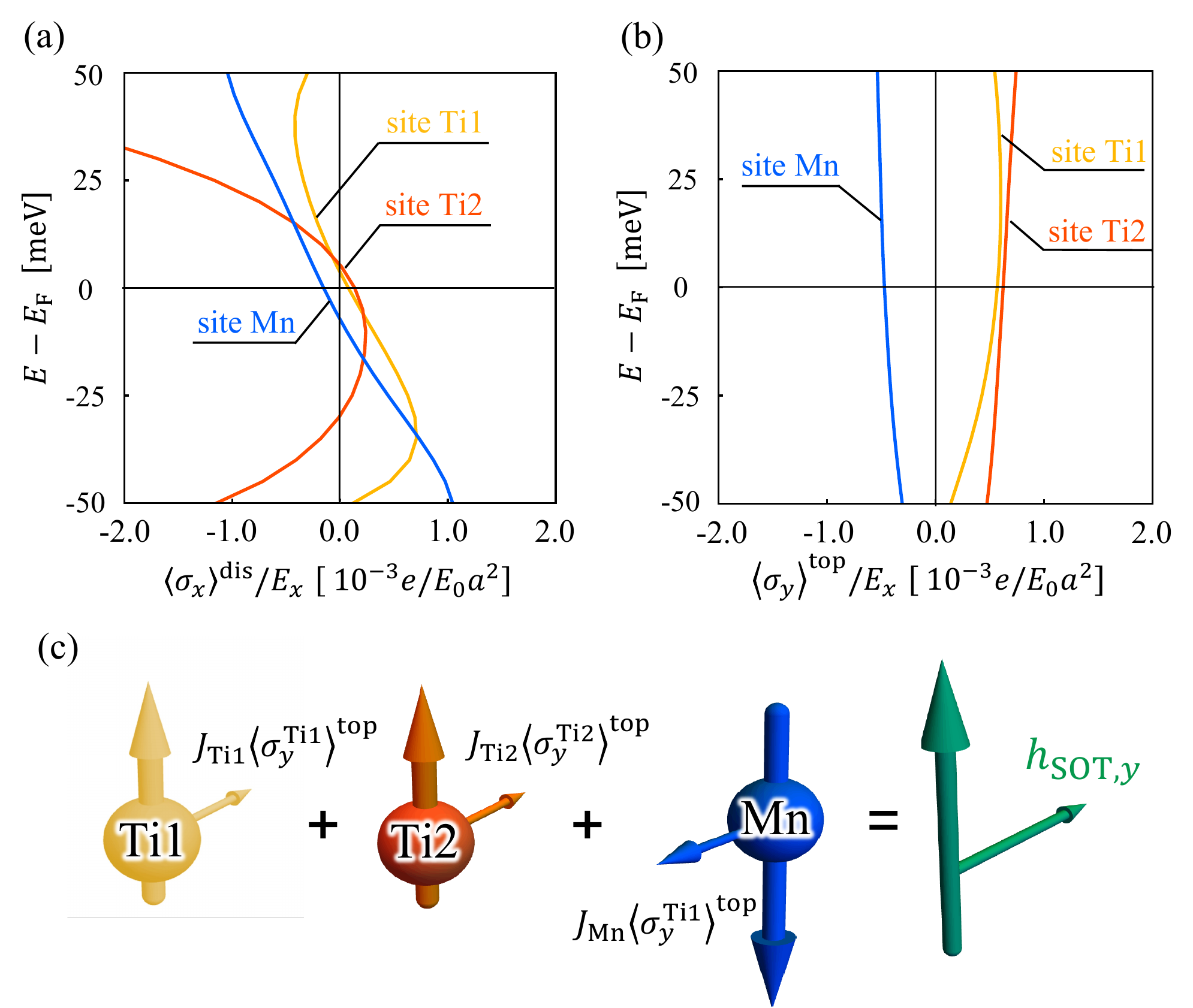}
\caption{
The energy dependences of (a) the dissipative and (b) topological spin densities on each sublattice when the electric field is applied to the $+x$ direction.
(c)
$\langle \sigma^{\alpha}_{y}\rangle^{\rm top}$ 
are schematically shown by the arrows on each sublattice 
$\alpha = \ce{Ti}1$~(yellow), 
$\ce{Ti}2$~(red), and $\ce{Mn}$~(blue). These spin densities are additive on the compensated ferrimagnetic ordering~(green) and act as an effective magnetic field~$\bm{h}_{\rm SOT}$.
} 
\label{fig:torque}
\end{figure}
%===== FIG =====%
%
%
%
To understand the origin of the electrically induced spin density,
we decompose Eq.~(\ref{eqn:SpinDens}) into two parts as
% $\sigma_{\alpha i}=\chi^{dis}_{\alpha i}+\chi^{top}_{\alpha i}$, where
%}
% S^{\alpha, \mu}_{nm,\bm{k}} J^{x}_{mn,\bm{k}} 
\begin{align}
% \begin{equation}\label{eqn:Spin_intra}
% \begin{split} 
 \frac{\braket{\sigma^{\alpha}_{\mu}}^{\rm dis}}{E_x}
 &= 2\tau\sum_{n}
    \int_{\bm{k}} 
     \frac{\partial f_{n\bm{k}}}{\partial E} 
     S^{\alpha, \mu}_{nn,\bm{k}} J^{x}_{nn,\bm{k}},\label{eqn:Spin_intra}\\
     % \langle n\bm{k}|  \frac{\hbar}{2}\hat{\sigma}^{\alpha}_{\mu} |n{\bm{k} \rangle} 
     % \langle n\bm{k}| (-e \hat{v}_{x})         |n{\bm{k} \rangle} 
% % \end{split}
% % \end{equation}
% \label{eqn:Spin_intra}
% \end{align}
% %
% %
% \begin{align}
% \begin{equation}\label{eqn:Spin_inter}
% \begin{split} 
 \frac{\braket{\sigma^{\alpha}_{\mu}}^{\rm top}}{E_{x}}
 &= -2\hbar \sum_{n \neq m}
    \int_{\bm{k}} f_{n\bm{k}}
     \frac{
      {\rm Im} [ S^{\alpha, \mu}_{nm,\bm{k}} J^{x}_{mn,\bm{k}}]
     }{
      (E_{n\bm{k}} - E_{m\bm{k}})^2 + (\frac{\hbar}{2\tau})^2
     }.
     % \frac{ \langle n\bm{k}|   \frac{\hbar}{2}\hat{\sigma}^{\alpha}_{\mu} |m{\bm{k} \rangle} 
     %         \langle m\bm{k}|   (-e   \hat{v}_{x} )   |n{\bm{k} \rangle} }
     % { (E_{n\bm{k}} - E_{m\bm{k}})^2 + (\frac{\hbar}{2\tau})^2}.
% \end{split}
% \end{equation}
\label{eqn:Spin_inter}
\end{align}
%  
%Here we use the relation, $\frac{1}{E_n-E_m+i\eta} =  -i\pi \frac{\eta / \pi}{(E_n-E_m)^2 + \eta^{2}} + \frac{(E_n-E_m)^2}{(E_n-E_m)^2 + \eta^{2}}$.
%Here we use the relation, $\frac{1}{x+i\eta} =  -i\pi \frac{\eta / \pi}{x^2 + \eta^{2}} + \frac{x^2}{x^2 + \eta^{2}}$ for $x = E_{n} - E_{m}$.
% \textcolor{blue}{
Equation (\ref{eqn:Spin_intra}) includes
% }
only the electronic states near $E\_F$ 
% \textcolor{blue}{
from $\partial f_{n\boldsymbol{k}}/\partial E$,
% } 
%the contribution of the Fermi surface.
%We call this term as {\it dissipative} response in the following. 
whereas Eq.~(\ref{eqn:Spin_inter}) describes the contribution from all the electronic states below $E\_F$ 
% \textcolor{blue}{
from $f_{n\boldsymbol{k}}$.
% }
% 
% \textcolor{blue}{
Thus, we classify
% }
$\braket{\sigma_\mu }^{\rm dis}$  as the contribution to the ``{\it dissipative} SOT'', and $\braket{\sigma_\mu}^{\rm top}$ as the ``{\it topological} SOT''.
We show later that $\braket{\sigma_\mu}^{\rm top}$ in our model is indeed described by the topological feature called the MBC.

%=====================================================

%=================2nd paragraph=======================
Based on the above formulae, 
we study the $E\_F$-dependence of the induced spin densities.
Figures~\ref{fig:torque}(a) and \ref{fig:torque}(b) respectively show the energy dependences of 
$\braket{\sigma_{\mu}}^{\rm dis}$ 
and 
$\braket{\sigma_{\mu}}^{\rm top}$ 
with a fixed relaxation time $\tau=3 \hbar/E_0$.
In Fig.~\ref{fig:torque}(a),
% one can find that
% \textcolor{blue}{
the magnitudes of
% }
$\braket{\sigma_{\mu}}^{\rm dis}$ on all the three sites are supressed around $E\_F$, namely, around the energy of the WPs.
Such an energy dependence is similar to the results obtained 
with the minimal Weyl Hamiltonian with the isotropic spin-momentum locking in continuum~\cite{DKurebayashi2021,Burkov2023}.
Contrary, as shown in Fig.~\ref{fig:torque}(b), 
$\braket{\sigma_{\mu}}^{\rm top}$ for all sublattices
are always nonzero around $E\_F$ and  nearly independent of $E$.
Such a topological response 
% \textcolor{blue}{
was not seen
% }
in the minimal Weyl Hamiltonian~\cite{DKurebayashi2021, Burkov2023},
% because the contribution from each WP vanishes due to
% \textcolor{blue}{
which is the artifact of
% }
the isotropic spin-momentum locking structure 
% \textcolor{blue}{
in the Fermi sea
% }
around each WP.
% 
% Remarkably, in our tight-binding model, such cancellation is violated,
% \textcolor{blue}{
Since the SOC structure and the WP distribution of \TMA considered here are anisotropic in momentum space,
it is capable of showing
% }
the finite topological response $\braket{\sigma_{\mu}}^{\rm top}$.
% and yielding a dominant contribution near the equilibrium Weyl point.
We
% \textcolor{blue}{
find that, in response to $\boldsymbol{E}$ applied to the $x$-direction,
$\langle \bm{\sigma}^\alpha\rangle^{\rm dis}$ points to the $x$-direction while $\langle \bm{\sigma}^\alpha\rangle^{\rm dis}$ to the $y$-direction.
% }
% checked that, as torque acting on magnetization $\bm n$, only $\braket{\sigma_{x}}^{\rm dis}$ and $\braket{\sigma_{y}}^{\rm top}$ are finite when the electric field is applied to the $x$-direction.
% In other situations, for example, when the electric field is applied to the $y$-direction,
% only $\braket{\sigma_{y}}^{\rm dis}$ and $\braket{\sigma_{x}}^{\rm top}$ are finite.

The topological response leads to the highly efficient conversion from current to spin density, because of the low current density.
To compare the spin density generated per unit current on each sublattice with the conventional Rashba system, 
we evaluate a dimensionless index 
$\zeta^{\alpha} = {e v\_F} \frac{\braket{\sigma^{\alpha}_{y}}^{\rm top}}{\braket{j_{x}}^{\rm dis}}$.
Here, $\braket{j_{x}}^{\rm dis}$ is the longitudinal current flowing in the $x$-direction, which is calculated by the formula similar to Eq.~(\ref{eqn:Spin_intra}), 
by replacing  $\hat{\sigma}_{\mu}$
with $-e \hat{v}_x$.
$v\_F = 6.3\times 10^4$ m/s is the Fermi velocity of the Weyl cone.
With $E = E\_F$, 
we find these dimensionless indices 
$\zeta^{\rm Ti1} = 5.6\times 10^3$, 
$\zeta^{\rm Ti2} = 5.0\times 10^3$, and 
$\zeta^{\rm Mn} = 4.2\times 10^3$.
These values are about $10^4$ times larger than the typical value $0.12$ in the two-dimensional Rashba system~(see Appendix B). %\ref{sec:2D_Rashba}).
In other words, in \TMA, we can reduce the current by the factor of $\sim 10^{-4}$ from that in the Rashba systems to induce the same magnitude of spin density.
% 
% 
% 
% 

%=====================================================

\begin{figure}[t]
   \includegraphics[width=1.0\hsize]{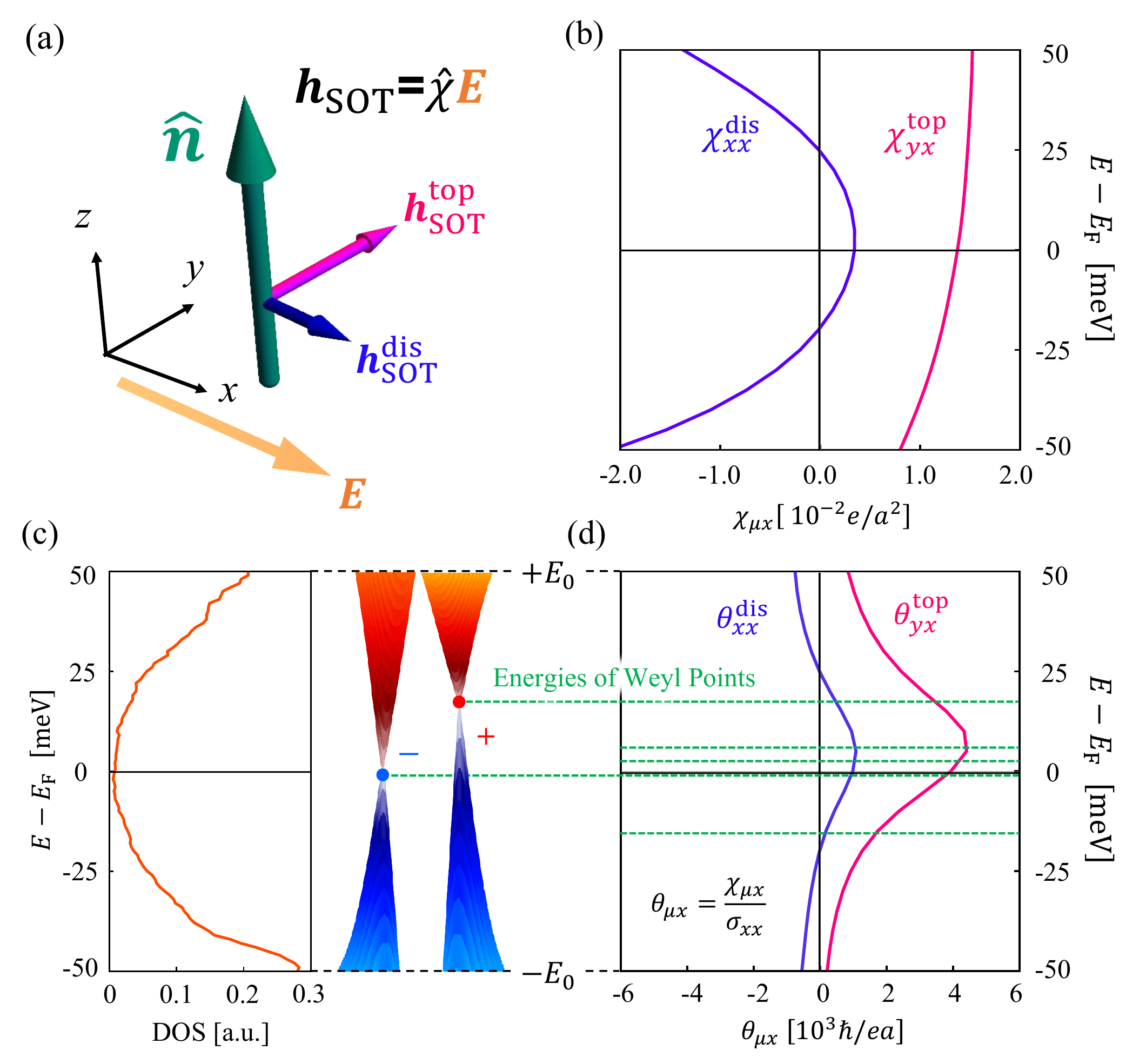}
\caption{
(a)~The effective magnetic fields form dissipative and topological response, acting on the compensated ferrimagnetic ordering $\hat{\bm{n}}$.
(b)~The energy dependences of the response coefficients $\chi^{\rm dis}_{xx}$ and $\chi^{\rm top}_{yx}$.
(c)~The density of states and 
one of the energy dispersion around WPs. 
The inversion symmetry breaking in \TMA causes energy differences between WPs.
(d)~The energy dependences of efficiencies of SOT $\theta^{\rm dis}_{xx}$ and $\theta^{\rm top}_{yx}$. 
} 
\label{fig:H_eff}
\end{figure}

%\section{Effective magnetic field and spin-orbit torque }
{\it Effective magnetic field and spin-orbit torque---}~
Above results show that, near $E\_F$, the signs of spin densities at Ti1 and Ti2 and that at Mn are opposite, as schematically shown in  Fig.~\ref{fig:torque}(c). 
These spin densities exert torques on the magnetic moments at the Ti sites and the Mn sites oppositely to each other.
During the dynamics of these magnetic moments,
violation of their compensation is predicted to be 
energetically prohibited \cite{Meguro2024}.
Thus, we treat
the compensated ferrimagnetic ordering $\hat{\bm{n}}$
as the only kinematical variable in this system.
We consider the effective magnetic field acting on $\hat{\bm{n}}$ in the linear response regime~\cite{Tserkovnyak2005} :
\begin{equation}\label{eqn:T_SOT}
  \begin{split} 
     \bm{h}_{\rm SOT}(\bm{r}) 
          &= \left\langle \frac{\partial {H}}{\partial \hat{\bm{n}} } \right\rangle  
          = \sum_{\alpha} 
             J_{\alpha} \langle \bm \sigma^{\alpha}(\bm{r}) \rangle.
  \end{split}
  \end{equation}
Here, $\braket{\bm{\sigma}^{\alpha}}$ is given by the formula in Eq.~\eqn{SpinDens}.
Note that $\bm{h}_{\rm SOT}$ has a unit of energy density. 
In later discussions, we will convert it to the unit of tesla and compare it with other systems.
% }
This $\bm{h}_{\rm SOT}$ exerts the SOT on $\hat{\bm{n}}$, expressed as 
$\bm{T}_{\rm SOT} = \hat{\bm{n}} \times \bm{h}_{\rm SOT}$.
Figure~\ref{fig:H_eff}(a) depicts the effective magnetic field for dissipative contribution $\bm{h}^{\rm dis}_{\rm SOT}$~(blue arrow) from $\braket{{\bm \sigma}^{\alpha}}^{\rm dis}$ and 
topological one $\bm{h}^{\rm top}_{\rm SOT}$~(red arrow) from $\braket{{\bm \sigma}^{\alpha}}^{\rm top}$.

%===== FIG =====%
%===== FIG =====%

% \textcolor{blue}{
First,
% }
we investigate the $E$ dependences of the effective magnetic field with a fixed relaxation time $\tau=3 \hbar/E_0$.
% \textcolor{blue}{
We characterize the
% }
effective magnetic field induced by the electric field $\bm{h}^{\rm dis/top}_{\rm SOT} = \hat{\chi}^{\rm dis/top} {\bm E}$
% \textcolor{blue}{
by the response coefficient tensor $\hat{\chi}^{\rm dis/top}$,
which is given by
% }
% 
%
\begin{align}%\label{eqn:}
  \chi^{\rm dis/top}_{\mu x} &= \sum_{\alpha} J_{\alpha} \frac{{\langle \sigma^{\alpha}_{\mu} \rangle}^{\rm dis/top}}{E_x}.
\end{align}
In Fig.~\ref{fig:H_eff}(b), the components of $\hat{\chi}^{\rm dis}$ and  $\hat{\chi}^{\rm top}$ are shown as functions of $E$.
Near $E\_F$, $\hat{\chi}^{\rm top}$ is larger than $\hat{\chi}^{\rm dis}$ and shows a relatively moderate energy dependence.
% We find that $\hat{\chi}^{\rm dis}$ is finite even near $E\_F$, 
% which does not appear in the continuum model.
%\textcolor{magenta}{This is because, owing to the sub-lattice structure of this model,}
%=================3rd paragraph=======================
To characterize the efficiency of the SOT per current, we introduce the following quantity, 
\begin{equation}\label{eqn:h_eff_efficiency}
  \theta^{\rm dis/top}_{\mu x} 
  = \frac{h_{{\rm SOT},\mu}^{\rm dis/top}}{\braket{j_{x}}^{\rm dis}}
  =  \frac{\chi^{\rm dis/top}_{\mu x}}{\sigma_{xx}}.
\end{equation}
Here, $\sigma_{xx}$ is a longitudical conductivity and has an minimum around the energy of the WPs, reflecting the structure of DOS shown in Fig.~\ref{fig:H_eff}(c).
Figure~\ref{fig:H_eff}(d) shows the energy dependence of $\theta^{\rm dis}_{xx}$ and $\theta^{\rm top}_{yx}$.
In Fig.~\ref{fig:H_eff}(d), one can find that 
$\theta^{\rm top}_{yx}$ has a peak near $E\_F$ and is dominant over $\theta^{\rm dis}_{xx}$.
The peak structure of $\theta^{\rm top}_{yx}$ arises from the combination of the large topological SOT and the small longitudinal conductivity around the energy of the WPs.
By measuring $\bm{h}_{\rm SOT}$ in the unit of Tesla
(see Appendix C),
we find that $\theta^{\rm top}_{yx}$ is of the order of 
$10^{-13}\ {\rm T m}^{2}/{\rm A}$
within the calculated energy range, reaching the value 
$\theta^{\rm top}_{yx} = 5.09 \times 10^{-13}\ {\rm T m}^{2}/{\rm A}$ at $E\_F$.
This value is 10-1000 times larger than the SOT efficiency measured in 
the heterostructures of ferromagnets and heavy metals \cite{Suzuki2011, Emori2013}, 
% and also those measured in
and also those calculated in
the bulk antiferromagnetic ${\rm Mn}_{2}{\rm Au}$~\cite{Zelezny2014,Zelezny2017} and $\rm CuMnAs$~\cite{Wadley2016}.
% }
Owing to the topological contribution,
% }
\TMA can generate an SOT with the efficiency 
much higher than those previous reports.

%===== FIG =====%
\begin{figure}[t]
   \centering
   \includegraphics[width=1.0\hsize]{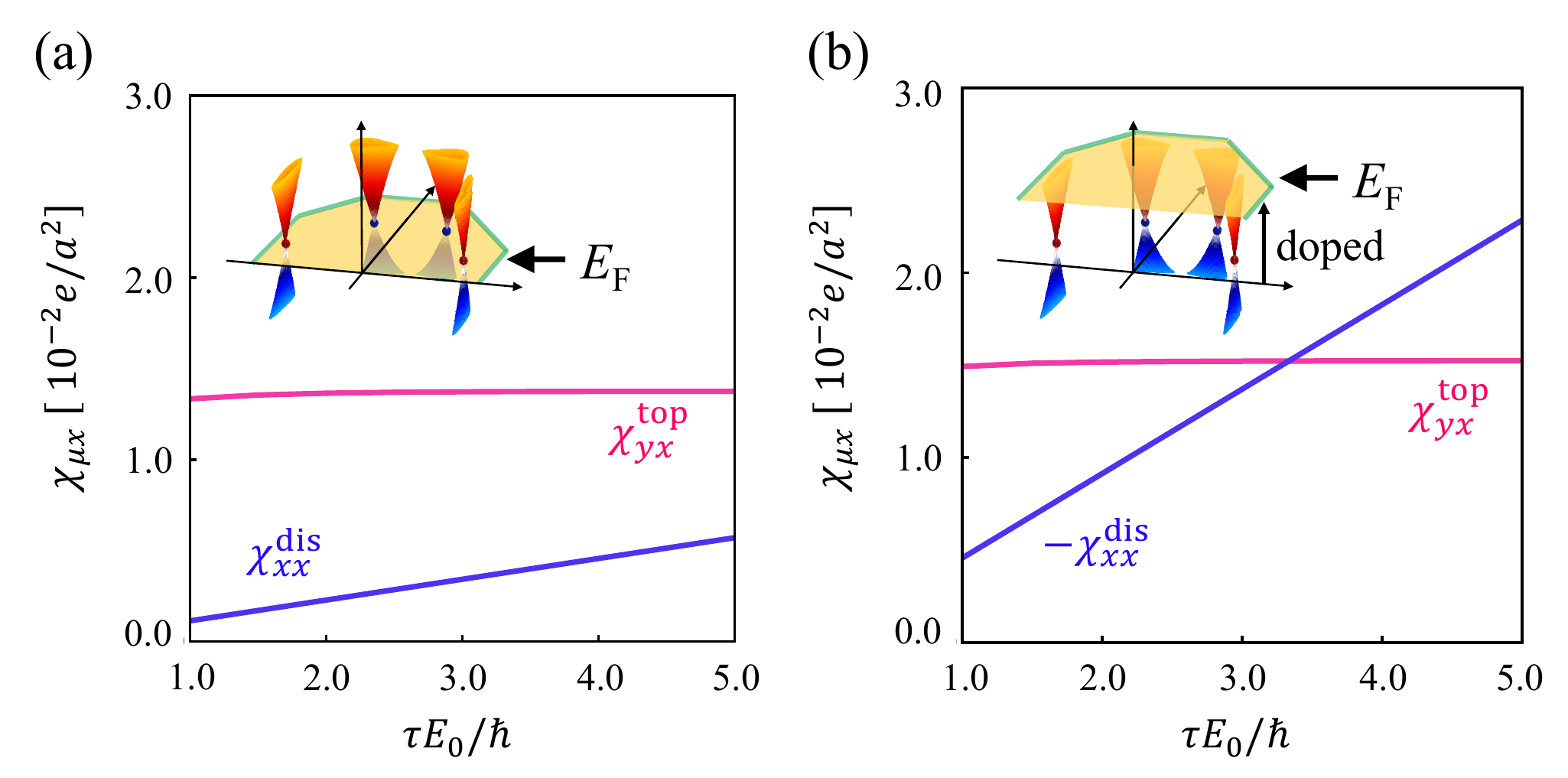}
\caption{
% Relaxation time dependences for response coefficients $\chi^{\rm dis}_{xx}$ and $\chi^{\rm top}_{yx}$.
% (a) When $E\_F$ is located at the energy of the WPs, 
% topological response $\chi^{\rm top}_{yx}$ is dominant 
% and almost independent on the relaxation times compared to $\chi^{\rm dis}_{xx}$.
% (b) When $E\_F$ is away from the energy of the WPs, dissipative response $\chi^{\rm dis}_{xx}$ can be dominant due to the larger fermi surface than (a).
% It also depends on the relaxation times strongly.
% 
Relaxation time dependences for response coefficients $\chi^{\rm dis}_{xx}$ and $\chi^{\rm top}_{yx}$ when 
(a) $E\_F$ is located at the energy of the WPs and 
(b) $E\_F$ is away from the energy of the WPs.
} 
\label{fig:tauDep}
\end{figure}
%===== FIG =====%

%-------------------------
Next, we consider the relaxation-time dependences of $\chi^{\rm dis}_{xx}$ and $\chi^{\rm top}_{yx}$ for non-doped and doped cases,
% \textcolor{blue}{
which are shown in Figs.~\ref{fig:tauDep}(a) and \ref{fig:tauDep}(b).
% }
% when $E = E\_F$ is located at the energy of the WPs.
% 
% In Fig.~\ref{fig:tauDep}(a) where $E\_F$ is set as the energy of the WPs, 
In the non-doped case 
with the $4/6$-filling,
% }
$E\_F$ is located near the energy of the WPs,
% \textcolor{blue}{
as shown in the inset of Fig.~\ref{fig:tauDep}(a).
% }
% 
% \textcolor{blue}{
From the calculation results in Fig.~\ref{fig:tauDep}(a),
we find
% }
that $\chi^{\rm top}_{yx}$ is almost independent of $\tau$, 
% \textcolor{blue}{
which implies the insensitivity
% }
to the disorder.
Moreover, throughout the entire range of $\tau$ we have calculated, 
$\chi^{\rm top}_{yx}$ remains larger than $\chi^{\rm dis}_{xx}$.
% 
% \textcolor{blue}{Such a} behavior of $\chi^{\rm top}_{yx}$ is related to the MBC from 
% the WPs, as will be discussed later.
% 
In contrast, $\chi^{\rm dis}_{xx}$ is proportional to $\tau$ as seen from Eq.~(\ref{eqn:Spin_intra}).
With regard to magnitude of $\chi^{\rm dis}_{xx}$, 
the DOS is small enough, and hence $\chi^{\rm dis}_{xx}$ becomes also small 
% \textcolor{blue}{
in comparison with
% }
$\chi^{\rm top}_{yx}$.

%-------------------------

% -------- paragraph 1 -----------

Then, we investigate the dependence of $\chi^{\rm dis}_{xx}$ and $\chi^{\rm top}_{yx}$ on $\tau$ in doped case where $E\_F$ is 50 meV above the energy of the WPs,
which are shown in Fig.~\ref{fig:tauDep}(b).
% 
% \textcolor{blue}{
We find
% }
that both the $\tau$-dependence and magnitude of $\chi^{\rm top}_{yx}$ are almost unchanged from those in Fig.~\ref{fig:tauDep}(a). 
In contrast, $\chi^{\rm dis}_{xx}$ is still proportional to $\tau$, while its magnitude becomes larger.
The magnitude of $\chi^{\rm dis}_{xx}$ becomes comparable to that of $\chi^{\rm top}_{yx}$ and exceeds it around $\tau = 3.5 \times \hbar/E_0$.
This is because the DOS becomes larger when $E\_F$ is shifted away from the energy of the WPs.
% 

% \textcolor{blue}{
From the discussions above, we find that
% }
$\chi^{\rm dis}_{xx}$ is always proprtional to $\tau$, with its magnitude dependent on $E\_F$, which is consistent with the previous study with the minimal Weyl Hamiltonian 
% \textcolor{blue}{
in continuum%
% }
~\cite{DKurebayashi2021}.
Our main finding in this study is that the topological response $\chi^{\rm top}_{yx}$ is present irrespective of $\tau$,
and is almost independent of the choice of $E\_F$.
% \textcolor{blue}{
Such a
% }
behavior of $\chi^{\rm top}_{yx}$ is related to the 
% \textcolor{blue}{
band geometry characterized by the
% }
MBC from the WPs, as 
% \textcolor{blue}{
we shall discuss in the following.
% }

%\section{Mixed Berry curvature and discussion}\label{sec:discussion}

%========== 1st paragraph ===================
{\it Mixed Berry curvature and discussion---}%
So far we have found 
the topological SOT $\chi^{\rm top}_{yx}$ is dominant compared to the dissipative SOT  $\chi^{\rm dis}_{xx}$ in our model of \TMA.
To understand the origin of this $\chi^{\rm top}_{yx}$, we study the MBC~\cite{Freimuth2014}.
It is a BC extended in the composite parameter space 
$(\bm{k}, \hat{\bm{n}})$ spanned by
the momentum $\bm{k}$ and the 
ferrimagnetic ordering $\hat{\bm{n}}$. 
The $yx$ components
of the BC $\Omega^{\bm{k}\bm{k}}_{yx}$ and 
the MBC $\Omega^{\hat{\bm{n}}\bm{k}}_{yx}$ for the occupied bands are expressed as
\begin{equation}
\begin{split} 
 \Omega^{\bm{k}\bm{k}}_{yx}(\bm{k}) 
  &= 2\,{\rm Im}\! \sum_{n \neq m}
     f_{n\bm{k}}  
     \frac{ 
      \langle
       n\bm{k}| \partial_{k_{y}}{H} |m\bm{k} 
      \rangle 
      \langle
       m\bm{k}| \partial_{k_{x}}{H} |n\bm{k}
      \rangle
     }
     {(E_{n\bm{k}} - E_{m\bm{k}})^2},
\\
 \Omega^{\hat{\bm{n}}\bm{k}}_{yx}(\bm{k}) 
  &= 2\,{\rm Im}\! \sum_{n \neq m}
     f_{n\bm{k}}  
     \frac{
      \langle
       n\bm{k}| \partial_{\hat{n}_{y}}{ H} |m\bm{k}
      \rangle 
      \langle
       m\bm{k}| \partial_{k_{x}}{ H} |n\bm{k}
      \rangle
     }
     {(E_{n\bm{k}} - E_{m\bm{k}})^2}.
\end{split}
\end{equation}
% 
%===== FIG =====%
\begin{figure}[t]
   \centering
   \includegraphics[width=1.0\hsize]{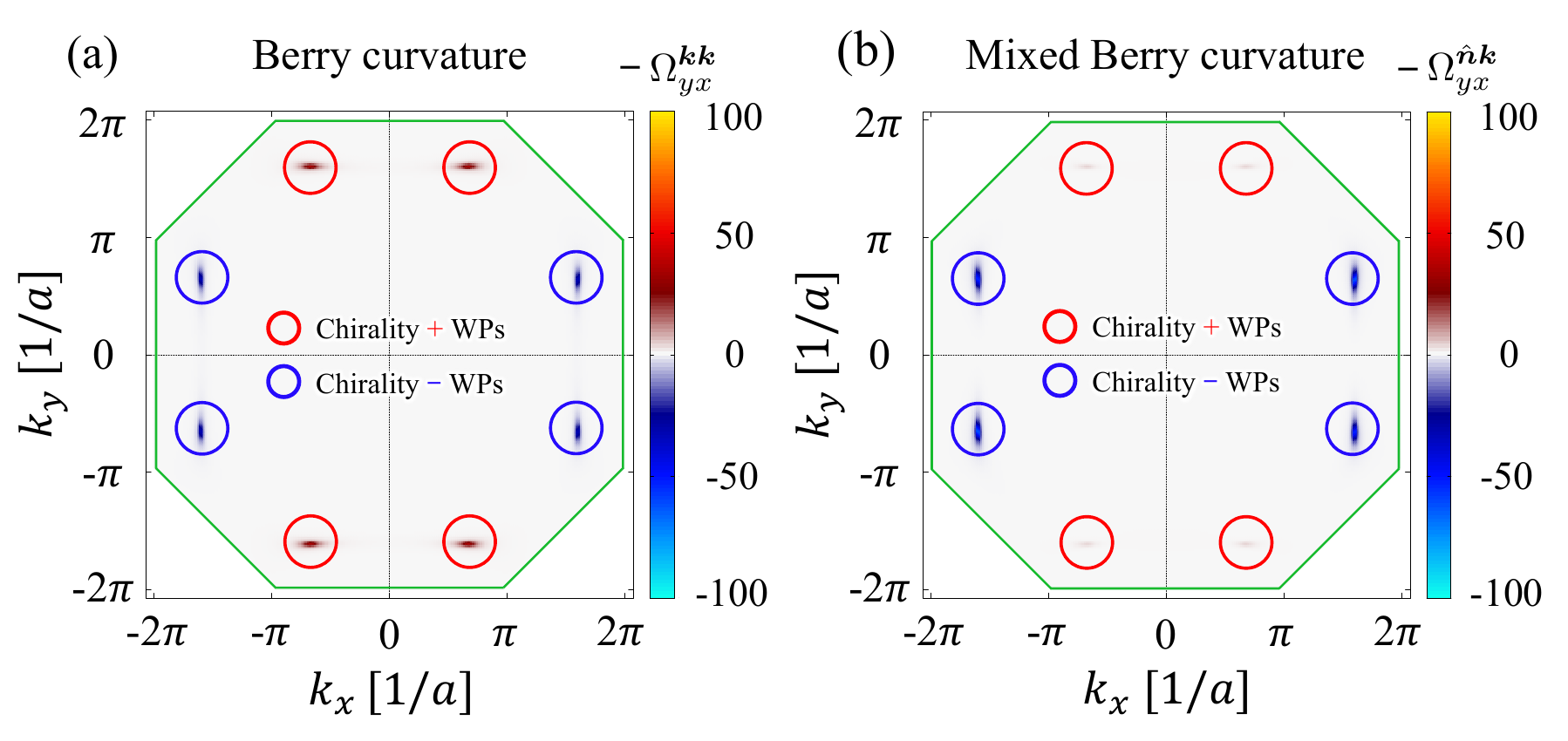}
\caption{
The mapping of the $yx$ component of 
(a)~the BC $-\Omega^{\bm{k}\bm{k}}_{yx}$ and 
(b)~the MBC 
$-\Omega^{\hat{\bm{n}}\bm{k}}_{yx}$ 
on $k_{z} = 0.1 [1/a]$ at $E\_F$.
In momentum space, the positions of the peaks/valleys of 
$-\Omega^{\bm{k}\bm{k}/\hat{\bm{n}}\bm{k}}_{yx}$ 
correspond to the positions of the WPs.
} 
\label{fig:mometumResolve}
\end{figure}
%===== FIG =====%
% 
The topological SOT [Eq.~(\ref{eqn:Spin_inter})] can be rewritten with the MBC~\cite{Freimuth2014,Manchon2019,xiao2021}, 
% F.Y.Chen2023 必要??
  \begin{equation}
     \bm{h}^{\rm top}_{\rm SOT} 
           %&= \left \langle \frac{\partial {\hat H}}{\partial \bm{n}} \right \rangle
           = e\hbar \int_{\bm k}  ~{\hat \Omega}^{\hat{\bm{n}}\bm{k}}(\bm{k})~{\bm E},
  \end{equation}
in the clean limit $\tau \rightarrow \infty$.

% \textcolor{blue}{
To understand the source of the MBC contributing to the large SOT found in \TMA,
we compare
% }
the distributions of the BC $-\Omega^{\bm{k}\bm{k}}_{yx}$
and the MBC $-\Omega^{\hat{\bm{n}}\bm{k}}_{yx}$
on the $k_{x}$-$k_{y}$ 
plane in Fig.~\ref{fig:mometumResolve}.
% 
%Recall that each $k_{i}$-$k_{j}$ plane contains the eight WPs.
We find that the MBC shows peak structures around the WPs,
%We have checked that the strengths of the MBC in the other momentum region are much smaller than those from the WPs.
% These peak structures are 
which are similar to those of the ordinary BC contributing to the intrinsic anomalous Hall effect.
We have checked that the distributions of the MBC on the other planes containing the WPs show similar peak structures.
% 
% \textcolor{blue}{
These numerical results suggest that, even
% \textcolor{blue}{
in the absence of spin-momentum locking%
% }
,
% though spin-momentum locking does not hold, 
% \textcolor{blue}{
the velocity operator
% }
$\frac{\partial {H}}{\partial \bm{k}}$ and 
% \textcolor{blue}{
the effective magnetic field operator
% }
$\frac{\partial {H}}{\partial \hat{\bm{n}}}$ have similar structures near the WPs,
yielding the peaks of both the ordinary and MBCs.
% }
As a result, the WPs serve as the origin not only of the intrinsic anomalous Hall effect but also of the topological SOT.
%============================================

% 
{\it Conclusion---}
Our study clarifies that the ferrimagnetic Weyl semimetal \TMA shows distinct behavior in the SOT.
The previous studies 
on the SOT in MWSMs discussed only the dissipative SOT, but not the topological SOT~\cite{DKurebayashi2021, Burkov2023}.
In contrast, the present study finds that ferrimagnetic Weyl semimetal \TMA
% uniquely 
shows the giant topological SOT, 
which is useful for dissipation-free switching of magnetization. 
The key element for this topological SOT is the MBC from all the occupied states.
In \TMA, this topological quantity is notably enhanced around the WPs in momentum space.
Moreover, due to the tiny carrier density around the energy of the WPs,
the longitudinal current gets suppressed.
% 
% Furthermore,
% in a quantitative manner,
As a result of these characteristics in \TMA,
the efficiency of the effective magnetic field per the injected current reaches around
% Our calculations yield a current-induced effective magnetic field of the efficiency
10-1000 times those estimated previously in noncentrosymmetric antiferromagnets.
% 
% This is owing to the MBC mentioned above and  
% the tiny carrier density around the energy of the WPs.
These results suggest that the design of the MBC 
in magnetic topological systems, such as MWSMs,
is important for the further development of SOT-based spintronics devices.

% Since the MMWSMs exhibit the....
% one expects that...
% With a specific model of MWSM, we theoretically demonstrated the efficient electrical spin generation by the MBC.

% that the MBC can be an advantage in controlling the magnetization dynamics. 

% in rea

%(practical? -> efficiency and device design) 
% From the practical point of view, 
% our results also suggested that the expected spin-torque efficiency is much larger than in conventional systems, 
% enabling us to control magnetization with a small current.
% In addition, compensated ferrimagnetic ordering without exhibiting no net stray field can be an advantage for designing highly integrated devices.
% These features extend the possibility of efficient and functional spintronic devices based on the magnetic Weyl semimetals. 
% }

%\acknowledgments
{\it Acknowledgments---}
%The authors would like to appreciate %T.~Misawa
% H.~Tsuchiura, and %  just in case, please ask Kentaro
% A.~Tsukazaki 
%for valuable discussions.
This work was supported by
JST CREST, Grant Nos.~JPMJCR18T2
and by
JSPS KAKENHI, Grant Nos.~%
JP20H01830,  % Kiban-B Nomura
JP22K03446, % Kiban-C Kobayashi
JP22K03538, % Kiban-C Araki
and
JP23K19194. % Startup, Ozawa
% A.~O.~was supported by
% JST CREST, Grant No.~JPMJCR19T3
% and 
% JP23K19194.
%Start 

%\begin{thebibliography}{10}

%\bibliography{ref}

\section*{End matter}

% \subsection{Details of the tight-binding model}\label{TMA_model_detail}
{\it Appendix.A : Details of the tight-binding model---}\label{TMA_model_detail}
We here explain the details of the hopping term and SOC term in our tight-binding model based on our previous study~\cite{Meguro2024}.
The primitive unit cell is the FCC type  and the unit vectors are 
$\bm{a}_{1} = \frac{a}{2}(1,1,0)$,
$\bm{a}_{2} = \frac{a}{2}(0,1,1)$ and
$\bm{a}_{3} = \frac{a}{2}(1,0,1)$, where $a$ is a lattice constant and set $a=1$ for simplicity.
In the model Hamiltonian [see Eq. (\ref{eqn:Ham0})],
the first term includes inter-sublattice hopping with strengths 
$t_{\ce{Ti}1-\ce{Ti}2}$, 
$t_{\ce{Ti}2-\ce{Mn}}$, 
and $t_{\ce{Mn}-\ce{Ti}1}$, 
and the intra-sublattice hopping with strength 
$t_{\ce{Ti}1-\ce{Ti}1}$,
$t_{\ce{Ti}2-\ce{Ti}2}$, and
$t_{\ce{Mn}-\ce{Mn}}$.
The second term is the on-site energy and $\epsilon_\alpha$ is its strength. 
%-----------------------------------------------%
%----------------- paragraph 4 ------------------%　　ここでスピン軌道相互作用
The third term is SOC and originates from the intrinsic electric field,
which is generated due to the local breaking of inversion symmetry. 
The dominant asymmetry comes from the imbalance between Ti1 and Al sites. 
We consider SOC for the Ti2-Ti2 and Mn-Mn hoppings with the same strength $\lambda_{\rm SOC}$,
and neglect the SOC for the Ti1-Ti1 for simplicity.
$\bm{d}^{\alpha ij}_{1,2}$ are the two nearest-neighbor hopping vectors from the site $i$ to $j$ of the sublattice $\alpha$,
which was originally introduced in a Fu-Kane-Mele model~\cite{Fu2007}. 
% \\
%----------------- paragraph 5 ------------------%ここで交換相互作用
The fourth term represents the exchange coupling between conduction electron spin and magnetic moment.
$\bm{\sigma}$ is the Pauli matrix vector representing electron spin.
$\hat{\bm{n}}$ is the unit vector representing the compensated ferrimagnetic ordering.
$J_\alpha$ is the coupling strength on sublattice $\alpha$.

The hopping parameters are set to 
$t_{\ce{Ti}1-\ce{Ti}2}=420$ meV,
$t_{\ce{Ti}2-\ce{Mn}}=176.4$ meV,
$t_{\ce{Mn}-\ce{Ti}1}=504$ meV,
$t_{\ce{Ti}1-\ce{Ti}1}=21$ meV,
$t_{\ce{Ti}2-\ce{Ti}2}=357$ meV, and
$t_{\ce{Mn}-\ce{Mn}}=-21$ meV.
 On-site energies 
$\epsilon_{\ce{Ti}1} = \epsilon_{\ce{Ti}2} = \epsilon_{\ce{Mn}} = -903$ meV.
 The strengths of the exchange coupling are 
$J_{\ce{Ti}1} = J_{\ce{Ti}2} = 294$ meV,
$J_{\ce{Mn}} =- 714$ meV.
The sign difference corresponds to the ferrimagnetic ordering.
 The strength of SOC is 
$\lambda_{\rm SOC} = -84$ meV.
%  We set the energy unit 
% $t_{0}= 420$ meV. 
The parameters are set so that the energy spectrum including the WPs,
the DOS, and the strength of the anomalous Hall conductivity 
become consistent with first-principles calculations\cite{Shi2018}. \\

{\it Appendix.B : Spin density and its efficiency of 2D Rashba Model---}\label{sec:2D_Rashba}
2D Rashba Model explains the two dimensional electron gas~(2DEG) with the Rashba type SOC.
Calculating the SOT by this model, only the $yx$($xy$) component of the dissipative contribution is finite. It is given by the following ~\cite{Edelstein1990,J.Inoue2003},
  \begin{equation}\label{eqn:Rashba_SOT}
  \begin{split} 
     \frac{\braket{\sigma_{y}}^{\rm dis}}{E_{x}} &= \frac{2 m^{*}e}{\hbar^2} \lambda \tau
  \end{split}
  \end{equation}
Here, $m^{*}$ is the effective mass of 2DEG, and $\lambda$ is the strength of Rashba type SOC.
With the electron density $n$, the DC conductivity is also given by 
$\frac{\braket{j_{x}}^{\rm dis}}{E_{x}} = \frac{ne^2}{m^{*}} \tau$.
% %
The dimensionless index $\zeta$ which explains spin density generated per unit current is given by, 
  \begin{equation}\label{eqn:Rashba_eff}
     \zeta = {e v\_F} \frac{\braket{\sigma_{y}}^{\rm dis}}{\braket{j_{x}}^{\rm dis}} 
     = \frac{2 m^{*2} \lambda v\_F}{\hbar^{2}n}
  \end{equation}
Here, as the effective mass~$m^{*}$ and electron density~$n$ of the 2DEG, the typical values of $0.068$$m_{e}$ and $10^{15}$~$\mathrm{m}^{-2}$ are chosen respectively.
The strength of SOC is $1.51 \times 10^{4}$~\cite{J.Nitta1997}. 
$v\_F$ is the Fermi velocity, given by $v\_F=\hbar k\_F/m^{*}$, where Fermi wavevector $k_{\rm F}$ is assumed $ 0.1 {\rm nm}^{-1}$. 
Substituting these typical values into Eq.~(\ref{eqn:Rashba_eff}), one can obtain a value $\zeta = 0.12$.\\

{\it Appendix.C : Estimation of effective magnetic field of ${\rm Ti}_{2}{\rm MnAl}$---}\label{sec:Heff_unit}
Here, we explain details of the estimation of the SOT efficiency per current as defined by Eq.~(\ref{eqn:h_eff_efficiency}).
First, we estimate the magnitude of the electric field in \TMA when a current is applied. 
Given a typical experimental current density of $j = 10^{10}\ {\rm A}/{\rm m}^2$ 
and the longitudinal conductivity of \TMA at $E\_F$, $\sigma_{xx} = 0.24 \frac{e^2}{a\hbar}$, 
the electric field in \TMA is estimated to be $10^{5}\ {\rm V}/{\rm m}$.
This value is used in the following estimation of the magnitude of the effective magnetic field.
% 

% For $\braket{\sigma^{\alpha}_{\mu}}^{\rm top}$, 
Referring to Fig.~\ref{fig:torque}(b), 
we approximate that the spin densities from all electronic states below $E_{\rm F}$ on each sublattice 
have almost the same magnitude : 
% the magnitude of the spin densities of $\rm Ti1$, $\rm Ti2$, and $\rm Mn$ are nearly equal at $E\_F$ : 
$
\braket{\bm{\sigma}^{\rm Ti1}}^{\rm top} \approx
\braket{\bm{\sigma}^{\rm Ti2}}^{\rm top} \approx
-\braket{\bm{\sigma}^{\rm Mn}}^{\rm top} 
\equiv
\bm{s}^{\rm top}
$.
With this assumption,
the effective magnetic field $\bm{h}_{\rm SOT}$ defined by Eq.~(\ref{eqn:T_SOT}) is rewritten as, 
\begin{align}\label{eqn:Huc_TMA}
 \bm{h}_{\rm SOT}^{\rm top}
  &= \sum_{\alpha = {\rm Ti}1,{\rm Ti}2,{\rm Mn}} 
       J_{\alpha} \langle \bm{\sigma}^{\alpha} \rangle^{\rm top} \nonumber\\
  &= \left(
       J\_{Ti1} \braket{\bm{\sigma}^{\rm Ti1}}^{\rm top}
      +J\_{Ti2} \braket{\bm{\sigma}^{\rm Ti2}}^{\rm top}
      +J\_{Mn}  \braket{\bm{\sigma}^{\rm Mn}}^{\rm top}
     \right) \nonumber\\
  &\approx \left( 
       J\_{Ti1} + J\_{Ti2} - J\_{Mn}
     \right) \bm{s}^{\rm top}.
\end{align}
We note that $\bm{h}_{\rm SOT}$ defined here has the dimension of energy density, not magnetic field.
To convert it to the dimension of magnetic field, tesla,
we need to divide it by the magnetic moment for each site,
\begin{equation}\label{eqn:Huc_TMA}
  \bm{B}^{\alpha}_{\rm SOT}
       = \frac{J_{\alpha}\frac{a^3}{2}}{m_{\alpha}} \braket{\bm{\sigma}^{\alpha}}.
\end{equation}
Here, $\frac{a^3}{2}$ represents the volume of the conventional unit cell of \TMA, and 
$m_{\alpha}$ denotes the magnitude of the magnetic moment residing on sublattice $\alpha$.
This $\bm{B}^{\alpha}_{\rm SOT}$ can be expressed with $\bm{h}_{\rm SOT}$ as, 
\begin{align}\label{eqn:Beff_henkei}
 \bm{B}^{\alpha,{\rm top}}_{\rm SOT}
   &= \frac{J_{\alpha}\frac{a^3}{2}}{m_{\alpha}} \langle {{\bm \sigma}}^{\alpha} \rangle^{\rm top} \nonumber\\
   &\approx \frac{1}{m_{\alpha}} J_{\alpha} \frac{a^3}{2} \bm{s}^{\rm top} \nonumber\\
   &= \frac{1}{m_{\alpha}} \frac{J_{\alpha}}{J\_{Ti1} + J\_{Ti2} - J\_{Mn} } \frac{a^3}{2}\bm{h}^{\rm top}_{\rm SOT}.
\end{align}
For instance, the field strength at the Ti1 site reads 
% \textcolor{blue}{
$|\bm{B}^{\rm Ti1, {\rm top}}_{\rm SOT}| \approx 1.15 \times 10^{-3} \ {\rm T}$%
% }
, by substituting into Eq.~(\ref{eqn:Beff_henkei}) the quantities
$m_{\rm Ti1} = 1.44 \mu_{B}$ \cite{Skaftouros2013}, 
$J_{\ce{Ti}1} = J_{\ce{Ti}2} = 294$ meV,
$J_{\ce{Mn}} =- 714$ meV, 
lattice constant $a = 6.24 \times 10^{-10}\ {\rm m}$ \cite{Skaftouros2013}, 
the electric field in \TMA $E_x = 10^{5} {\rm V}/{\rm m}$, and 
the response coefficient at $E\_F$, 
% \textcolor{blue}{
$\chi^{\rm top}_{yx} = 1.36 \times 10^{-2}\frac{e}{a^2}$
% }
obtained from Fig.~\ref{fig:H_eff}(b).
As discussed in the main text, the spin densities of Ti1 and Ti2 have signs opposite to that of Mn. 
Therefore, the effective magnetic fields acting on the compensated ferrimagnetic order $\bm n$ are additive across the sublattices [see Fig.~\ref{fig:torque}(c)]. 
By calculating $\bm{B}^{\alpha,{\rm top}}_{\rm SOT}$ for Ti2 and Mn and summing them, the magnetic field acting on the compensated ferrimagnetic ordering $\bm n$ is found to be 
% \textcolor{blue}{
$\bm{B}^{\bm{n}}_{\rm SOT} \approx 5.09 \times 10^{-3}\ {\rm T}$
% }
, or
% \textcolor{blue}{
$\theta^{\rm top}_{yx} = B^{\bm{n}}_{{\rm SOT},y}/j_{x} = 5.09 \times 10^{-13}\ {\rm T~m^{2}}/{\rm A}$
% }
with respect to the longitudinal current $j_x$.

%\input{main_1st_revised}
%\bibliography{main}
%\bibliographystyle{apsrev}

\end{document}